\documentclass[aps,prl,preprintnumbers,amsmath,amssymb,latexsym,nofootinbib,array,enumerate,letter,twocolumn]{revtex4-1}

\pdfoutput=1

\usepackage{graphicx}
\usepackage{color}
\definecolor{Mahogany}{rgb}{0.62,0.24,0.15}
\definecolor{DarkRed}{rgb}{0.6,0,0}
\definecolor{DarkGreen}{rgb}{0,0.6,0}
\definecolor{DarkBlue}{rgb}{0,0,0.6}
\definecolor{purple}{rgb}{0.6,0,0.8}
\usepackage[colorlinks=true,linktocpage=true,linkcolor=DarkBlue,citecolor=DarkGreen,urlcolor=DarkRed]{hyperref}

\usepackage{graphicx}
\usepackage{array,booktabs}
\usepackage{float}
\usepackage{amsmath}
\usepackage{multirow} 
\usepackage{lineno}
\usepackage{comment}

\usepackage[hang,flushmargin]{footmisc}
\usepackage{placeins}
\usepackage{bbold}
\usepackage{multirow}
\usepackage{slashed}
\usepackage{cancel}
\usepackage[caption=false]{subfig}

\newcommand*{\INFNFR}{Istituto Nazionale di Fisica Nucleare, Laboratori Nazionali di Frascati, C.P. 13, 00044 Frascati, Italy}

\begin{document}

\title{Probing light mediators at the MUonE experiment}
\author{Giovanni Grilli di Cortona}
\email{grillidc@lnf.infn.it}
\author{Enrico Nardi}
\email{enrico.nardi@lnf.infn.it}
\affiliation{\INFNFR}

\date{\today}

\begin{abstract}
The MUonE experiment, that aims to provide a precise measurement of the hadronic vacuum polarization contribution to the muon $g-2$ via elastic muon-electron scattering, has also the potential to explore the parameter space of light new physics. Exploiting the process $\mu^- N \to \mu^- N X$, where  $N$ is the target nucleus and X is a new physics light mediator, we demonstrate that MUonE can be sensitive to new regions of parameter space for sub-GeV dark photons. In particular, thanks to its muon beam, 
MUonE will be able to explore uncharted parameter space regions 
for the $L_\mu-L_\tau$ model. Finally, we also find that MUonE can probe the parameter space of axion-like particles for different assumptions of the couplings to electrons, muons and photons.
\end{abstract}

\maketitle

{\it Introduction}---%
Although the Standard Model (SM) has proven to be an extremely successful theory, there is a plethora of evidence that it is incomplete. Among others, the evidence for the existence of dark matter (DM) renders the idea of the presence of dark sectors very attractive. DM may interact with the SM via a mediator particle, with its mass and couplings spanning many order of magnitude. While heavy mediators are more and more constrained, the  
interest in light mediators with different characteristics (spin, couplings, masses) 
is steadily increasing. The corresponding signatures can be revealed  
with experiments covering a broad range of different energies. In particular, if the mediator is long-lived, it can be detected exploiting the displaced vertex  or the missing energy and momentum signatures. 
Among this class of new physics models we can list dark photons~\cite{Okun:1982xi,Galison:1983pa, Holdom:1985ag}, vector bosons with a gauged lepton flavor symmetry $L_\mu-L_\tau$~\cite{Foot:1990mn, He:1990pn,  Ma:2001md} and axion-like particles (ALPs)~\cite{Jaeckel:2010ni,Arvanitaki:2009fg,Irastorza:2018dyq}.

In this Letter, we investigate the potential reach of the MUonE experiment~\cite{CarloniCalame:2015obs,Abbiendi:2016xup,Abbiendi:2677471,Marconi:2019rio,Venanzoni:2018ktr,Abbiendi:2020sxw,Abbiendi:2022oks} to 
long lived mediators of sub-GeV mass that couple to muons and electrons. The MUonE experiment aim to measure the hadronic vacuum polarization (HVP) contribution to the anomalous muon magnetic moment $(g-2)_\mu$ exploiting the elastic $\mu-e$ scattering. This measurement will be crucial to shed light  on the current 
status  of the $(g-2)_\mu$ measurements~\cite{Muong-2:2006rrc,Muong-2:2021ojo}, 
which exhibit a $4.2\,\sigma$ discrepancy with the SM theoretical prediction 
when the HVP contribution is estimated via the R-ratio technique~\cite{Aoyama:2020ynm,Davier:2010nc,Davier:2017zfy,Davier:2019can}
while, if instead the HVP value is taken from  the current most precise available lattice determination~\cite{Borsanyi:2020mff},  the tension is substantially reduced to about $1.5\sigma$~\cite{Colangelo:2022jxc}
(see however Ref.~\cite{Darme:2021huc} for a way to reconcile this 
discordance by invoking indirect new physics effects). 
 
MUonE will collide high energy muons onto the atomic electrons of its beryllium (Be) or carbon (C) targets, measuring the electrons and muons final states with great precision. The high resolution tracking system~\cite{Abbiendi:2677471,Ballerini:2019zkk} allows the experiment to be very sensitive to displaced vertex signatures. 
Besides the determination of the HVP,  the experiment can also be able to perform   
searches for certain types of NP signals, in which case, as we will discuss in the following,  
$\mu$ scattering off  Be or C target nuclei rather than electrons would be the 
most favorable channel. 
In this Letter, we demonstrate that the MUonE experiment can probe dark photons or ALPs produced in the $\mu^- \mathrm{Be}(\mathrm{C}) \to \mu^- \mathrm{Be}\,(\mathrm{C})\, X$ process and decaying into electrons or muons. The final state will be characterized by three leptons, $\mu^-e^+e^-(\mu^-\mu^+\mu^-)$, where the $e^+e^-(\mu^+\mu^-)$ pair
originating from $X$ decays can be reconstructed to a vertex displaced from the target. New physics at MUonE was studied in~\cite{Dev:2020drf,Masiero:2020vxk,Asai:2021wzx,Galon:2022xcl}, and muon beams have been proposed to search for light mediators in~\cite{Chen:2017awl,Kahn:2018cqs,Chen:2018vkr,Gninenko:2019qiv,Galon:2019owl,Cesarotti:2022ttv}.

{\it The MUonE experiment}---%
The strategy motivating the MUonE experiment has been described in~\cite{Abbiendi:2677471}. The leading HVP contribution to the $(g-2)_\mu$ can be 
extracted from a very precise measurement of the differential cross-section for the process $\mu^\pm e^- \to \mu^\pm e^-$~\cite{CarloniCalame:2015obs}, from which one can infer,   
 after subtracting the   purely leptonic part that is theoretically know with high precision, the hadronic contribution to 
 the effective electromagnetic coupling in the space-like region. 
This measurement can be carried out by colliding muons from the CERN M2 muon beam (with an energy of $E_1=150-160$ GeV) off atomic electrons of the Be or C targets inside one of the forty consecutive and identical aligned modules of 1\,m length. Each module contains a $1.5$ cm thick target and three pairs of tracking square layers with an active area of $100$ cm$^2$. In order to aid in the identification and selection of the final state leptons, the forty modules are followed by an electromagnetic calorimeter (ECAL) and a muon detector.
The ECAL is able to resolve the muon-electron ambiguity for energies of the outgoing final states of $\mathcal{O}$(1) GeV, while the muon detector is used mainly to reduce the pion contamination.
MUonE anticipates a resolution in the longitudinal $z$ direction of about 0.1 cm, and it can efficiently detect displaced vertices for angles between tracks larger than 0.1 mrad~\cite{Galon:2022xcl}.

In order to detect the final state leptons originating from  $X$ decays, the decay has to occur  in the region between the target and the first tracking layer,
located at a distance of 15 cm. Given the 1\,mm longitudinal resolution, 
we  define conservatively a fiducial decay interval corresponding to the  
range $2\leq z/{\rm cm} \leq 14.5$, that is in between the target and the first tracker 
of each module.

{\it The models}---%
In what follows we discuss the new physics scenarios of dark photons, $L_\mu-L_\tau$ vector bosons, and ALPs. 
The dark photon is a new vector boson charged under a $U(1)'$ gauge symmetry and coupling to a lepton current:
\begin{eqnarray}
    \mathcal{L} &\supset& -\frac{1}{4} F'_{\mu\nu} F'^{\mu\nu} +\frac{1}{2}m_{A'}^2 A'^{\mu}A'_{\mu} \nonumber\\
    &-& i g A'_\mu \sum_{i} \left(Q_{\ell_i} \bar{\ell_i} \gamma^\mu \ell_i + Q_{\nu_\ell} \bar{\nu}_\ell \gamma^\mu P_L \nu_\ell \right),
\end{eqnarray}
where $\ell= e,\,\mu,\,\tau$ are the lepton fields, $A'_\mu$ is the dark photon field and $F'_{\mu\nu}$ its field-strength. In this Letter, we discuss dark photons ($Q_\ell=1$, $Q_{\nu_\ell}=0$, $g=\epsilon e$) and the gauged flavour symmetry $L_\mu-L_\tau$ ($Q_{\mu,\tau} = Q_{\nu_{\mu,\tau}} = \pm 1$). In the latter case, couplings between the dark photon and $\nu_\mu$ or $\nu_\tau$ are also present.

The dominant production mechanism happens when the incoming muon exchanges a virtual photon $\gamma^*$ with a nucleon in the target and radiates a dark photon via the bremsstrahlung process~\cite{Kim:1973he, Tsai:1973py, Tsai:1986tx, Bjorken:2009mm}. In the Weizsaecker-Williams approximation~\cite{vonWeizsacker:1934nji, Williams:1934ad}, the full scattering cross section is approximated by the $2\to2$ process $\mu \gamma^* \to \mu A'$, weighted by the effective photon flux. The differential cross section as a function of the fraction of energy carried by the dark photon $x=E_{A'}/E_1$ is~\cite{Kim:1973he, Tsai:1973py}
\begin{equation}
    \frac{d\sigma}{dx} = \frac{8 \alpha^3 \epsilon^2 \beta_{A'}}{(m_{A'}^2\frac{1-x}{x}+m_\mu^2 x)}\left(1-x+\frac{x^2}{3}\right)\chi,
\end{equation}
where $\beta_{A'}=\sqrt{1-m_{A'}^2/E_{A'}^2}$. The effective photon flux is given by~\cite{Kim:1973he, Tsai:1973py, Bjorken:2009mm, Jodlowski:2019ycu}
\begin{eqnarray}
    \chi&=&\int_{t_{\mathrm{min}}}^{t_{\mathrm{max}}}dt \frac{t-t_{\mathrm{min}}}{t^2}\biggl[\left(\frac{a^2t}{1+a^2t} \right)^2 \left(\frac{1}{1+t/d}\right)^2 Z^2\nonumber\\
    &+&\left(\frac{a'^2t}{1+a'^2t} \right)^2\left(\frac{1+t(\mu_p^2-1)/(4m_p^2)}{(1+t/(0.71\,\mathrm{GeV}^2))^4} \right)Z    \biggr],
    \label{eq:chi}
\end{eqnarray}
where $t_\mathrm{min}\simeq (m_{A'}^2/(2E_{A'}))^2$ and $t_{max} \simeq m_{A'}^2+m_{\mu}^2$. The first term in the square brackets parameterizes the elastic atomic form factor with $a=106\,Z^{-1/3}/m_e\,\,(111\,Z^{-1/3}/m_e)$ for Be (C) targets, and the nuclear form factor with $d=0.164\,A^{-2/3}$ GeV$^2$~\cite{Tsai:1973py}. The second term describes the inelastic atomic form factor, with $a'=571.4\,Z^{-2/3}/m_e\,\,(773\,Z^{-2/3}/m_e)$ for Be (C) targets, and the inelastic nuclear form factor, where $m_p=0.938$ GeV is the proton mass and $\mu_p=2.79$~\cite{Tsai:1973py} 
is the nuclear magnetic dipole moment.

The decay width of a dark photon to massive leptons is given by
\begin{equation}
\Gamma_{\ell^+\ell^-} = \frac{\epsilon^2\alpha}{3}m_{A'}\left(1+\frac{2 m_\ell^2}{m_{A'}^2}\right)\sqrt{1-\frac{4 m_\ell^2}{m_{A'}^2}}, 
\end{equation}
leading to a decay length of $d_{A'}\sim\mathcal{O}(10)$ cm for characteristic dark photon momenta and masses of $\mathcal{O}(10)$ MeV. 
The $L_\mu-L_\tau$ gauge boson can furthermore decay into $\nu_\mu$ or $\nu_\tau$ with a decay width $\Gamma_{\nu\nu} = g^2 m_{A'}/(24\pi)$. We have also taken into account dark photon hadronic decays.

The spontaneous breaking of a global symmetry at some large new physics scale naturally produces weakly coupled ALPs. Following an effective field theory approach, we focus only on the ALP interactions with leptons and photons after  electroweak symmetry breaking:
\begin{equation}
    \mathcal{L} \supset 
    \frac{1}{4}g_{a\gamma} a F_{\mu\nu} \tilde{F}^{\mu\nu} +\frac{1}{2}  (\partial_\mu a) \sum_{i} g_{a\ell_i} \bar{\ell_i} \gamma^\mu \gamma^5 \ell_i,
    \label{eq:ALPLag}
\end{equation}
where $a$ is the ALP field, 
the dual electromagnetic field strength tensor is defined as $\tilde{F}^{\mu\nu} = \epsilon^{\mu\nu\alpha\beta} F_{\alpha \beta}/2$ with $\epsilon^{0123}=-1$ and, 
for brevity, we have omitted writing the usual ALP kinetic and mass terms.   We  assume that the couplings $g_{a\gamma}$, $g_{a\ell}$ and the ALP mass $m_a$ are independent parameters. Furthermore, we take possible  couplings to quarks and gluons to be negligible compared to $g_{a\gamma}$ and $g_{a\ell}$. 

Analogously to the dark photon case, the dominant production mechanism is given by the bremsstrahlung process. The differential cross section for the process $\mu^- + \mathrm{Be}(\mathrm{C}) \to \mu^- +  \mathrm{Be}(\mathrm{C}) + a$ as a function of $x=E_a/E_1$ is given by~\cite{Tsai:1986tx}
\begin{equation}
    \frac{d\sigma}{dx} = 2 \alpha_a r_0^2 x \beta_{a}\frac{(1+2f/3) }{(1+f)^2}\chi,
\end{equation}
where $r_0=\alpha/m_\mu$, $f=m_a^2(1-x)/(m_\mu^2x^2)$ and $\alpha_a=g_{a\ell}^2/4\pi$. The effective photon flux $\chi$ is given by Eq. \eqref{eq:chi}.

From the Lagrangian  in Eq. \eqref{eq:ALPLag} we can derive the ALP decay rates to photons and leptons
\begin{equation}
    \Gamma_{\gamma\gamma} = \frac{g_{a\gamma}^2m_a^3}{64\pi}, \qquad 
    \Gamma_{\ell^+\ell^-} = \frac{g_{a\ell}^2}{8\pi}m_\ell^2 m_a\sqrt{1-\frac{4 m_\ell^2}{m_a^2}}. 
    \label{eq:ALPdecrate}
\end{equation}

{\it Search strategy}---%
The differential number of signal events per energy fraction $x$ for both models is given by 
\begin{equation}
    \frac{dN}{dx} = \mathcal{L} \, \frac{d\sigma}{dx} \, \mathcal{P}_\mathrm{dec}(x) \mathrm{BR}(X\to\ell^+\ell^-),
    \label{eq:dN/dx}
\end{equation}
where $X=A'$ or $a$, $\mathcal{L}=1.5 \cdot 10^{4}$ pb$^{-1}$ is the expected integrated luminosity of the MUonE experiment, and 
\begin{equation}
\mathcal{P}_\mathrm{dec}(x)~=~e^{-z_{\mathrm{min}}/d_{A'}(x)}-e^{-z_{\mathrm{max}}/d_{A'}(x)}
\end{equation}
is the probability for the dark photon to decay within the fiducial interval between $z_{\mathrm{min}}=2$ cm and $z_{\mathrm{max}}=14.5$ cm after the beginning of each module. Finally, we multiply the number of events obtained in one module, for the total number of identical modules $N_{\mathrm{mod}}=40$.

In this Letter, we study events in which the three pairs of tracker layers show three charged tracks, one from the initial muon beam and two from the decay products of the dark photon or the ALP. As a consequence, the angular acceptance depends only on the decay location relative to the target. Given the high energy of the incoming muon, its angular distribution is always within the angular acceptance of each module. The same is valid for the new physics particle produced. The production is dominated by $\theta_X\lesssim\mathrm{max}(m_X/E_1,m_\mu/E_1)$ (for larger angles the cross section decreases as $\theta_X^{-4}$), and the high energy of the collision implies that the new state $X$ is highly boosted and well between the angular acceptance. Finally, the leptons from its decay will also maintain the same direction. 

Our search strategy  relies on requiring a charged lepton pair, reconstructed to a displaced vertex, to pass through the three pairs of tracking layers. 
Due to the high beam energy, 
the final state muon from the beam  automatically passes this requirement. 
The condition that the decay products pass through the tracking layers, on the other hand, depends on the probability that the new  particle decays inside the fiducial volume defined for each module,  and on the laboratory frame opening angle of the lepton pair coming from the reconstructed displaced vertex $\theta_{\ell\ell} \simeq 2 m_X / E_X$. 
This angle is bounded from below due to the fact that we need the first tracking layer to resolve the two tracks, and from above because we want both the decay leptons to pass through the last tracking layer. 
Hence we require $0.001 < \theta_{\ell\ell} < 0.05$. 
We furthermore require that the energy of the final muon satisfies $E_\mu>5$ GeV while  that of $X$ must satisfy $E_X>10$ GeV. This  ensures that all the leptons in the final state have  energy $E_\ell \gtrsim \mathcal{O}(1)$ GeV. %
Imposing these cuts restricts the range of integration of Eq. \eqref{eq:dN/dx} to the region 
\begin{equation}
      \max\left(\frac{E_X^{\mathrm{min}}}{E_1},\frac{2 m_X }{ E_1 \theta_{\ell\ell}^{\mathrm{max}}}\right)<x<\min\left(\frac{2 m_X}{E_1 \theta_{\ell\ell}^{\mathrm{min}}},1-\frac{E_\mu}{E_1}\right).  \nonumber
\end{equation}
The above cuts suffice to illustrate the new physics reach 
of MUonE. Clearly, the experimental search strategy 
 can be refined and optimized, but this is beyond the scope of this Letter~\cite{Galon:2022xcl}.

{\it Backgrounds}---%
There are two kinds of background that can potentially affect the displaced vertex search. The first one is characterized by SM processes that could also yield displaced decays. Long lived particles, like neutral Kaons, could be produced in coherent or deep-inelastic $\mu$-nucleus scattering. In the coherent case the nucleus would only slightly recoil, leading to a soft hadronic emission. On the other hand, the deep-inelastic scattering may lead to hard emission together with additional radiation. This kind of signature should be identifiable and different from the displaced topology. 
The second category includes SM processes which are prompt but are mis-identified as displaced because of tracker inefficiencies. This can happen in processes where a virtual photon decaying into a lepton pair is produced or in Bethe-Heitler trident reactions~\cite{Bjorken:2009mm}.  

In the following, we assume that the background can be identified and properly subtracted by a dedicated analysis. 

\begin{figure}[t]
 \includegraphics[width=0.48\textwidth]{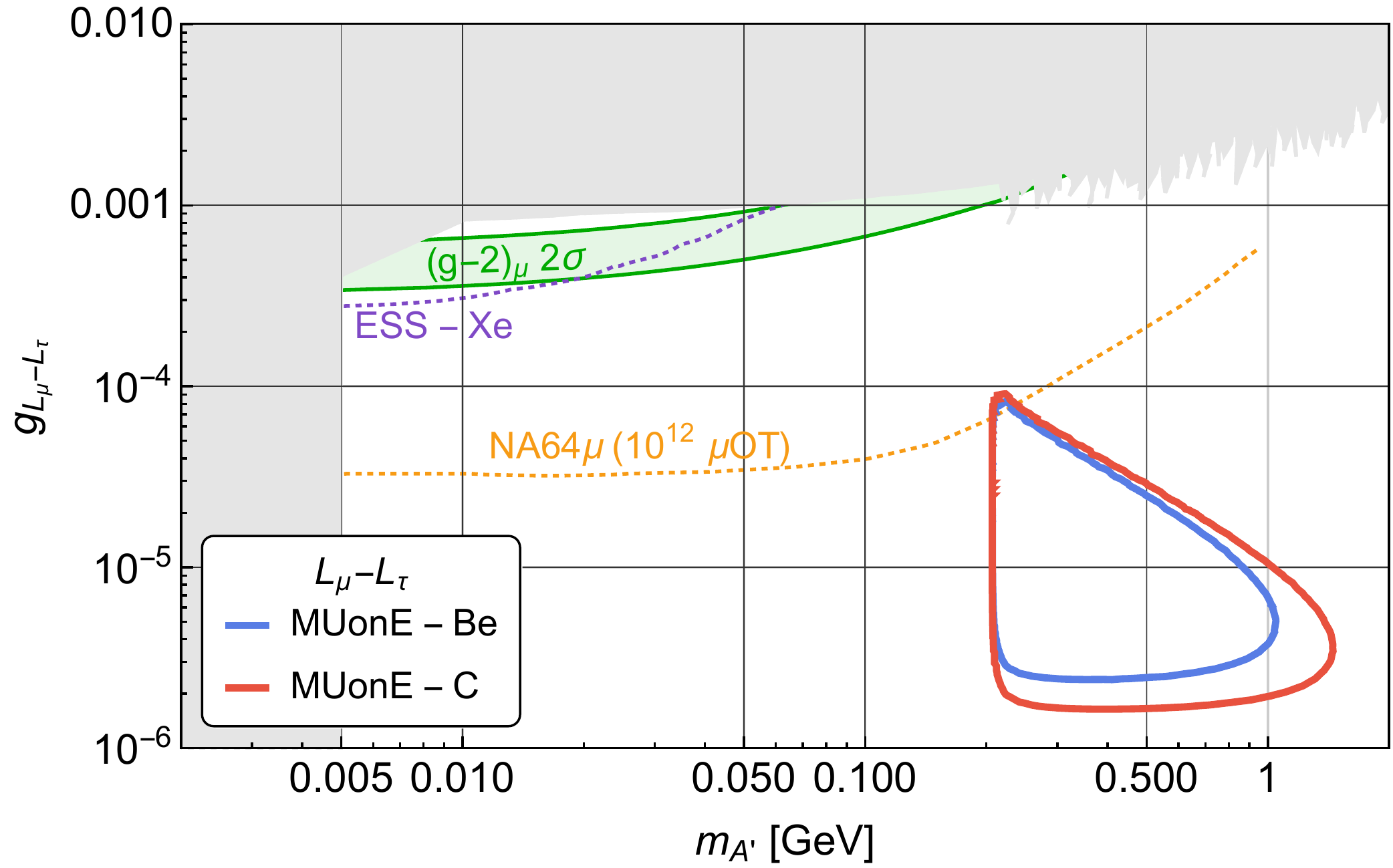}
 \caption{Future sensitivity at $90\%$ C.L. for the $L_\mu-L_\tau$ model at MUonE with a carbon (solid red) or beryllium (solid blue) target. Gray shaded regions are excluded by~\cite{BaBar:2016sci,Altmannshofer:2014pba, CCFR:1991lpl,Escudero:2019gzq,Gninenko:2020xys,Bellini:2011rx}. Dotted curves show future sensitivity from~\cite{Sieber:2021fue,Bertuzzo:2021opb}. We also show the $(g-2)_\mu$ $2\sigma$ preferred region~\cite{Muong-2:2006rrc,Aoyama:2020ynm,Muong-2:2021ojo,Davier:2010nc,Davier:2017zfy,Davier:2019can}.}
\label{fig:Lmu-Ltau}
\end{figure}

\begin{figure}[t]
 \includegraphics[width=0.48\textwidth]{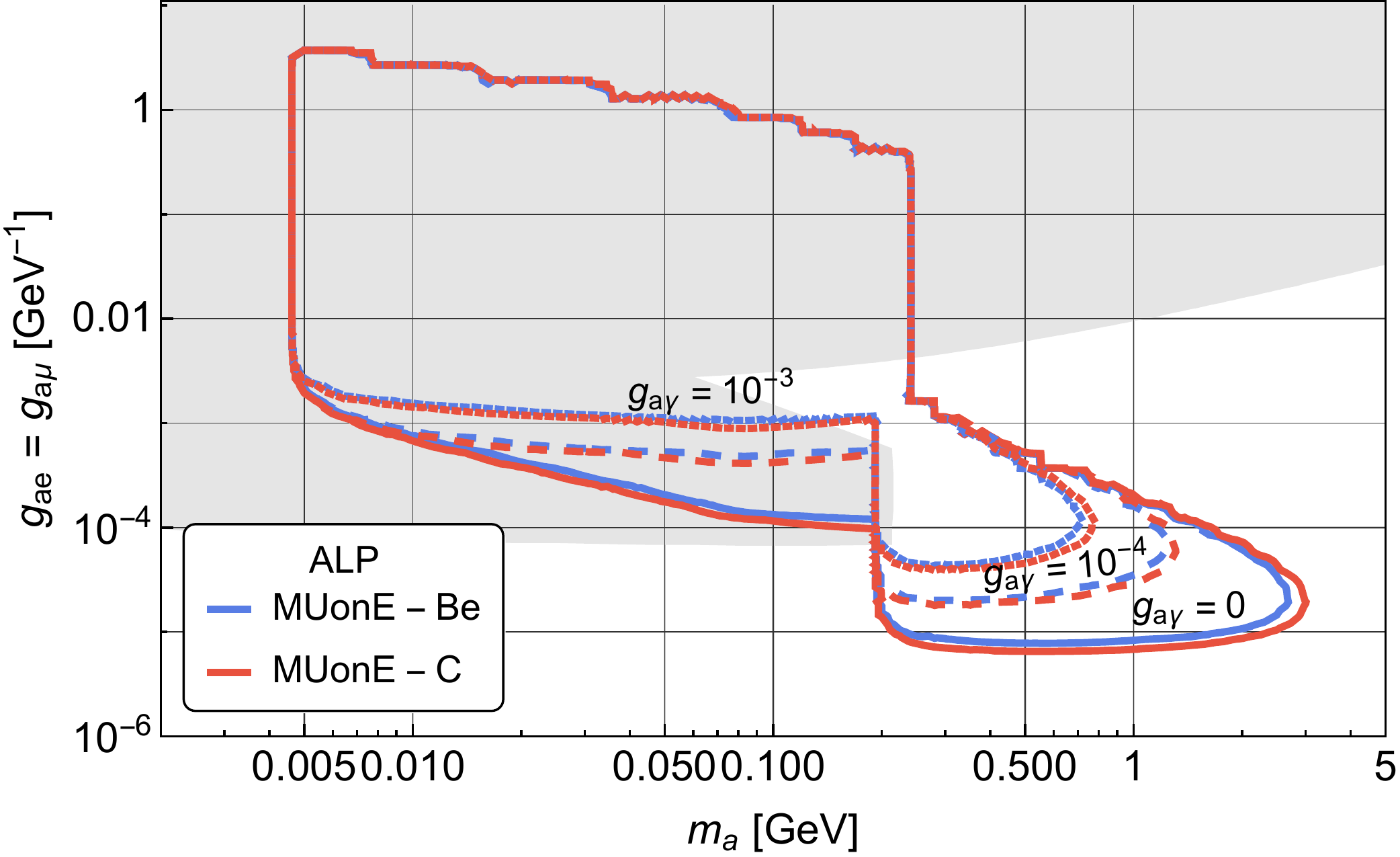}
 \caption{Future sensitivity at $90\%$ C.L. for ALPs at MUonE with a carbon (solid red) or beryllium (solid blue) target. Gray shaded regions are excluded by recasting~\cite{Bjorken:1988as,Essig:2010gu} and by the $(g-2)_\mu$ measurement \cite{Muong-2:2006rrc,Muong-2:2021ojo,Aoyama:2020ynm}. Dashed (dotted) colored curves show the sensitivity of MUonE for $g_{a\gamma}=10^{-4}$ $(g_{a\gamma}~=~10^{-3})$.}
\label{fig:ALP}
\end{figure}

{\it Results}---%
\begin{figure}[t]
 \includegraphics[width=0.48\textwidth]{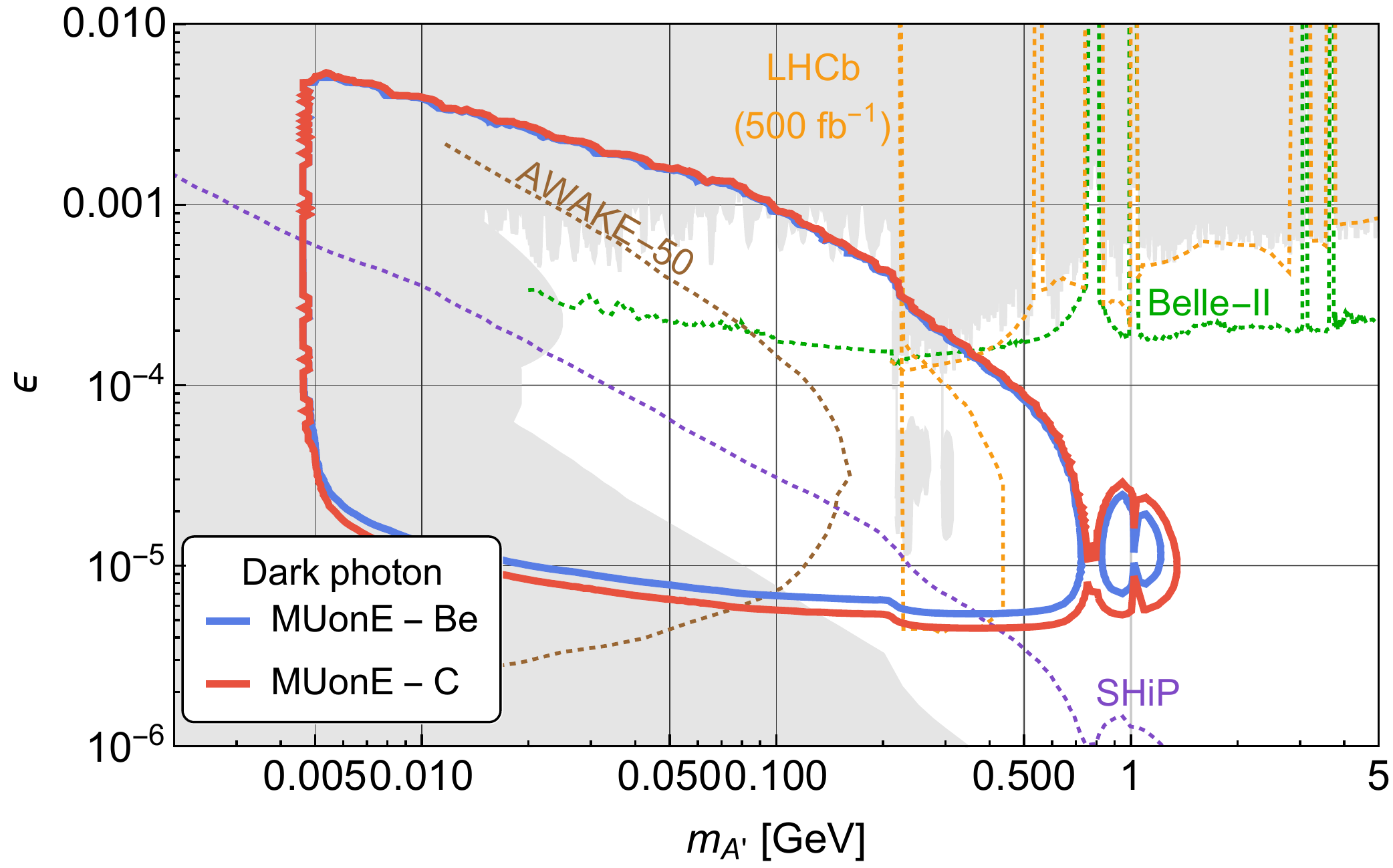}
 \caption{Future sensitivity at $90\%$ C.L. for the dark photon model at MUonE with a carbon (solid red) or beryllium (solid blue) target. Gray shaded regions are excluded, see Refs.~\cite{Riordan:1987aw,Bjorken:1988as, Batell:2014mga, Marsicano:2018krp,Bross:1989mp,CHARM:1985anb, Gninenko:2012eq,Blumlein:1990ay, Blumlein:2011mv, Blumlein:2013cua,BaBar:2014zli,Na482:2015wmo,Merkel:2014avp,KLOE-2:2011hhj, KLOE-2:2012lii, KLOE-2:2014qxg, KLOE-2:2016ydq,LHCb:2019vmc,Chang:2016ntp}. Dotted curves show future sensitivity as given in Refs.~\cite{Alekhin:2015byh,Ilten:2015hya, Ilten:2016tkc,Belle-II:2018jsg,Caldwell:2018atq}.}
\label{fig:dark_photon}
\end{figure} 
The reach of the dark photon, $L_\mu-L_\tau$ and ALP models at MUonE are presented in Figs.~\ref{fig:dark_photon},~\ref{fig:Lmu-Ltau} and~\ref{fig:ALP}. We consider an energy of the incoming muon of $E_1~=~160$~GeV for both a Be and a C target and we show exclusion curves at $90\%$ C.L.. 

The dark photon scenario in Fig.~\ref{fig:dark_photon} presents 
 in gray the existing bounds from beam dump experiments~\cite{Riordan:1987aw,Bjorken:1988as, Batell:2014mga, Marsicano:2018krp,Bross:1989mp,CHARM:1985anb, Gninenko:2012eq,Blumlein:1990ay, Blumlein:2011mv, Blumlein:2013cua}, from lepton pair resonance searches~\cite{BaBar:2014zli,Na482:2015wmo,Merkel:2014avp,KLOE-2:2011hhj, KLOE-2:2012lii, KLOE-2:2014qxg, KLOE-2:2016ydq,LHCb:2019vmc} and from the Supernova 1987A~\cite{Chang:2016ntp}. Furthermore, dotted curves denote the projected reach from SHiP~\cite{Alekhin:2015byh} (purple), LHCb~\cite{Ilten:2015hya, Ilten:2016tkc} (orange), Belle-II~\cite{Belle-II:2018jsg} (green), and AWAKE~\cite{Caldwell:2018atq} (brown). 

\bigskip

The reach of the MUonE experiment is shown as a blue (red) solid contour for a Be (C) target. 
The exclusion region is bounded at small dark photon masses due to the requirement of a minimum opening angle of the dark photon decay products, needed to resolve the two tracks in the first layer. For large couplings or large masses, the $A'$ decays before the decay region, while for small couplings it decays after the first tracking layer. As a consequence, the number of events is exponentially suppressed by $\mathcal{P}_\mathrm{dec}$. 
The slight kink around the muon mass threshold is due to the increased particle  width due to the opening of the decay channel $A'\to \mu^+ \mu^-$, which allows for decays in the fiducial volume for smaller values of $\epsilon$. 
The features in the limit at $m_{A'}\gtrsim 800$ MeV arise because of resonant production in the $A'\to \mathrm{hadrons}$ decay channel. 
These results may be modified by the presence of higher dimension operators \cite{Barducci:2021egn}.

The $L_\mu-L_\tau$ model is poorly constrained for $A'$  masses above $10^{-2}$ GeV and couplings below $10^{-3}$. 
The strongest constraints for this model come from searches for $A'$ decays at BaBar~\cite{BaBar:2016sci}, from neutrino trident production~\cite{Altmannshofer:2014pba, CCFR:1991lpl}, from measurements of the light nuclei primordial abundances~\cite{Escudero:2019gzq} and from a reinterpretation by~\cite{Gninenko:2020xys} of the Borexino limits~\cite{Bellini:2011rx}, shown in gray in Fig.~\ref{fig:Lmu-Ltau}. 
We furthermore show the $2\sigma$ region needed to explain the anomalous magnetic moment of the muon (green)~\cite{Muong-2:2006rrc,Aoyama:2020ynm,Muong-2:2021ojo,Davier:2010nc,Davier:2017zfy,Davier:2019can}, and the projected limits from NA64$\mu$~\cite{Sieber:2021fue} and from coherent elastic neutrino nucleus scattering searches at proposed detectors at the European Spallation Source~\cite{Bertuzzo:2021opb}. 

The sensitivity for a Be (blue) or C (red) target is bounded at low $m_{A'}$ by the muon production threshold. While the reach for this model is weak, complementary part of the parameter space could be probed at high energy muon beam dump experiments~\cite{Cesarotti:2022ttv}.

The projected sensitivity of MUonE for an ALP $a$ is shown in Fig.~\ref{fig:ALP} as a blue (red) curve for a Be (C) target. 
The simplified ALP model that we have discussed in this Letter has four free parameters: $g_{ae}$, $g_{a\mu}$, $g_{a\gamma}$ and $m_a$. A thorough investigation of the 4D parameter space of this model is beyond the scope of this work, so here we simply fix $g_{ae}=g_{a\mu}$ and $g_{a\gamma}=0$.

Analogously to the dark photon case, MUonE is not sensitive to masses below $m_a \simeq 4$ MeV, due to the angular acceptance for the lepton pair decay. 
At the muon threshold, assuming $g_{a\gamma} = 0$, the total decay width is dominated by the muon channel, as a consequence of the proportionality to the lepton mass squared. 
This means that once the muon decay is open, the ALP decays predominantly to a muon pair and its decay length at fixed $g_{ae}=g_{a\mu}$ is smaller, leading to vetoed ALP decays. 
For $m_a\gtrsim 3$ GeV, the ALP is too much long-lived and decays after the first tracking layer. 
We furthermore show how the sensitivity varies increasing the ALP coupling to photons: the dashed curves describe the sensitivity for $g_{a\gamma}=10^{-4}$, while the dotted one for $g_{a\gamma}=10^{-3}$. 
The gray shaded regions are excluded by a recasting of the experimental results by the E137~\cite{Bjorken:1988as}, following Ref.~\cite{Essig:2010gu}. 
This limit will become weaker if we assume smaller couplings to $g_{ae}$ or larger couplings to $g_{a\gamma}$. 
We also show in gray the limit from the experimental results on the $(g-2)_{\mu}$. 
Since the ALP-electron contribution is negative, we show the conservative bound obtained by taking the $5\sigma$ lower limit in~\cite{Muong-2:2006rrc,Muong-2:2021ojo,Aoyama:2020ynm}, such that $\delta a_\mu^{ALP} \geq 4.4\times 10^{-10}$. 
We computed $\delta a_\mu$ including the one loop contribution from Barr-Zee diagrams and the two-loop contribution from Light-by-Light diagrams, following the results by~\cite{Chang:2000ii,Marciano:2016yhf} and assuming a heavy new physics scale of $1$ TeV. 
A non vanishing coupling to photons gives a positive contribution to $a_\mu$, making the limit stronger. 

{\it Conclusions}---%
The current discrepancy between the experimental measurement of the muon anomalous magnetic moment and its SM computation could be resolved by an independent measurement of the hadronic-vacuum-polarization contribution. This is the main goal of the proposed experiment MUonE. 
In this Letter, we have shown that the characteristics of the experimental apparatus can also allow to search for light mediators coupled to electrons and muons. 
We have discussed which could be the sensitivity reach of MUonE for dark photons, $L_\mu-L_\tau$ gauge bosons and ALPs, relying on  the $\mu^- N \to \mu^- N X$ process, where $N=[$Be, C$]$ is the target material and $X=A'$ or $a$. 
This channel shows a larger potential reach with respect to the $\mu^- e^- \to \mu^- e^- X$ process \cite{Galon:2022xcl}, due to the coherent enhancement of the production cross section. 
The improvement in sensitivity with respect to existing experimental constraints and other proposed experiments is mainly due to the large energy of the muon beam and to the direct coupling to muons. The larger beam energy implies a larger boost factor. 
As a consequence it extends the decay length that can be probed, leading to the exploration of larger couplings and masses.

We have demonstrated that MUonE can have an excellent sensitivity to dark photon models for masses in the range $5\lesssim m_{A'}/\mathrm{MeV} \lesssim 10^3$ and 
couplings in the range $10^{-5} \lesssim \epsilon \lesssim 10^{-3}$, covering regions that will not be accessible to other forthcoming experiments like Belle-II, LHCb, SHiP, and AWAKE-50. 
Furthermore, MuonE can also explore a small region of parameter space for the $L_\mu-L_\tau$ model beyond the reach of NA$64\mu$. 
Finally, MUonE will be able to probe ALPS with couplings down to $g_{ae}=g_{a\mu}\simeq 10^{-5}$, for masses between the muon threshold and $m_a\simeq 3$ GeV. 
It would 
certainly be interesting to study more in detail how the sensitivity to ALP models 
would be modified by relaxing the assumptions on the couplings or by adding other invisible channels \cite{Darme:2020sjf}.

This  work presents a proof-of-concept analysis, in which we have identified the dominant backgrounds, arguing that they are negligible, and we have estimated the number of events for the signal. More accurate Monte Carlo simulations are of course needed. This Letter aims to further motivate the MUonE proposal, showing its potential to explore new physics beyond the $(g-2)_\mu$ problem. 

\vspace{0.2cm}

\begin{acknowledgements}
\noindent
{\it Acknowledgements}---%
This work has received support by the INFN Iniziativa Specifica Theoretical
Astroparticle Physics (TAsP). G.G.d.C. is supported by the Frascati National 
Laboratories (LNF) through a Cabibbo Fellowship, call 2019.
\end{acknowledgements}

\bibliography{biblio}
\end{document}